# Path Loss Models Based on Stochastic Rays[*]

L.-Q Hu*,

School of Electronic Science and Engineering, National University of Defense Technology, Changsha 410073, China, lqhu@ieee.org, lqhu@njupt.edu.cn

H. Yu,

School of Computer, Nanjing University of Posts and Telecommunications, Nanjing 210003, China

Y. Chen,

School of Electrical and Electronic Engineering, Nanyang Technological University, Singapore

Abstract: In this paper, two-dimensional percolation lattices are applied to describe wireless propagation environment, and stochastic rays are employed to model the trajectories of radio waves. We first derive the probability that a stochastic ray undergoes certain number of collisions at a specific spatial location. Three classes of stochastic rays with different constraint conditions are considered: stochastic rays of random walks, and generic stochastic rays with two different anomalous levels. Subsequently, we obtain the closed-form formulation of mean received power of radio waves under non line-of-sight conditions for each class of stochastic ray. Specifically, the determination of model parameters and the effects of lattice structures on the path loss are investigated. The theoretical results are validated by comparison with experimental data.

## 1    Introduction

In principle, radio wave propagation is governed by Maxwell equations, which provide the theoretical basis for the conventional wave-based approaches for wireless channel modeling [1]. Due to the complexity and sensitivity of implementing the wave-based models, the no-wave-based approaches have also been considered in practical electromagnetic engineering [2, 3, 4]. The concept of no-wave approaches first appeared at the end of the 1990's. The authors of [4] used the Boltzmann equation and diffusion equation to model propagation process of radio waves. The method based on the ideas in statistical physics is one of the initial attempts to model radio waves probabilistically and statistically. Subsequently, this concept was applied in [5] to construct a scattering environment, which provided a physical picture for understanding the importance of scattering in the transfer of information through wireless communication medium. The authors of [2] proposed the idea of modeling the urban propagation environment as a random lattice, where percolation theory was exploited in the domain of channel modeling. The authors of [6, 7] extended the work of [2] by modeling the radiating wave as a stream of photons propagating in the environment,

---

and the authors of [8] obtained the probability of stochastic rays under the constraints of the Manhattan distance. The trajectories of ultra-wideband multipath components are modeled as the samples of a Brownian bridge process in [9], where the stochastic rays are utilized. Moreover, the transport theory is also applied to analyze the path loss in indoor channels [10], which can be regarded as another kind of no-wave approaches by the aid of stochastic rays. A path loss model which features an exponentially-decaying signal power with the distance is obtained in [4, 10].

The distribution functions of generic stochastic rays under the constraints of the *Euclidean* propagation distance are obtained by using the no-wave approaches [11], where some the preliminary results on mean received power have been derived. Euclidean distance could be a more applicable propagation metric under certain circumstances due to the following reasons. Firstly, a probability density function (pdf) of rays of the number of collisions is generally met in electromagnetic engineering, where a receiver moves continuously along a path line. Secondly, the random lattice channels, or the site percolation, are homogeneous propagation environments. Especially in the case of sparse propagation scenario, the Euclidean distance of the transmitter and the receiver is natural. Thirdly, it has been shown through Monte Carlo counting procedure that, near the transmitting source the Euclidean symmetry dominates the scene [8]. Fourthly, the Euclidean distance is used pervasively in electromagnetic engineering application, and the determination of the Manhattan distance needs more information about the propagation scenario. In this paper we use site percolation lattices [12] to describe two-dimensional (2D) topography of propagation scenarios. The probability that a stochastic ray undergoes certain number of reflections at a specific spatial location is derived, which leads to a closed-form expression of mean received power of radio waves and the corresponding path loss models. The paper is organized as follows. Section 2 provides the theoretical formulation of stochastic rays. We construct random lattice channels based on the site percolation theory and derive the pdf of stochastic rays that undergo certain number of collisions at each location. Two topological parameters of propagation environment are defined. In Section 3, a rigorous derivation of mean received power and signal path loss is discussed. In Section 4, the experimental data are used to validate the theoretical results presented in Section 3. Finally, some concluding remarks are made in Section 5.

## 2  Theoretical model of stochastic rays
### 2.1  Random lattice channels and stochastic rays
To form site percolation, we may declare each vertex of a lattice to be open with probability *p* or closed otherwise, independent of all other vertices [12]. In the following discussion, we consider a 2D space $\mathbb{Z}^2$, where $\mathbb{Z}=\{...-1, 0, 1,...\}$ are integers. Moreover, if an interval of lattices or the cell-side length, *a*, is specified, then $a\mathbb{Z} \times a\mathbb{Z}$ defines a 2D space. There exists a critical value, namely the percolation threshold, $p_c \approx 0.59275$ [12]. For $p<p_c$, all the empty clusters are of finite size. On the other hand, for all $p>p_c$, an infinite empty cluster appears [12]. When *a*=2, *p*=0.3, 0.5, 0.7 and 0.9, respectively, the corresponding site percolation lattices are depicted in Fig. 1.

There are not so many differences from the site percolation lattices and the urban build-up areas. This fact has been pointed out in [2]. Moreover, these lattices also can be used to describe the indoor multipath channel where the propagation environment is highly cluttered.

The channels described by the site percolation are called random lattice channels. Generally the probability $p$ of lattices for modeling realistic urban or indoor region should be larger than $p_c$.

The radio waves propagating in random lattice channels form stochastic rays. This idea is very useful for wireless channel modeling. The multipath components (MPCs) are defined statistically and each stochastic ray can be considered as a sample of a stochastic process. Fig. 2 shows stochastic rays propagating in random lattice channels, where the rays are generated by Brownian bridge processes [9]. The positions of the transmitter and the receiver are the fixed positions of the Brownian bridge process. Only those sample trajectories with reflection points situating on the surfaces of closed vertex are regarded as effective paths.

*2.2 Distribution of stochastic ray and the average traveled distance*

Without loss of generality, let the transmitter be at (0, 0), and the receiver at ($x$, $y$). Given ($x$, $y$), the probability that arrived ray undergoing $i$ reflections among all the rays is denoted as $Q_i(x, y)$. Its analytic expressions can be obtained from maximum entropy approaches.

*Lemma 1:* (Maximum entropy distributions [13]) A probability density function $P(x, y)$ satisfies: (i) $P(x,y) \geq 0, x,y \in \mathbb{R}$; (ii) $\iint_{x,y} P(x,y) \mathrm{d}x \mathrm{d}y = 1$; (iii) $\iint_{x,y} \rho_i(x,y) P(x,y) \mathrm{d}x \mathrm{d}y = \alpha_i$. If the Shannon entropy is of the form

$$H(P) = -\iint_{x,y} P(x,y) \lg[P(x,y)] \mathrm{d}x \mathrm{d}y \tag{1}$$

the pdf $Q(x, y)$, which uniquely maximize $H(P)$ over all probability densities $P(x, y)$, satisfies

$$Q(x,y) = c e^{\eta \rho_i(x,y)} \tag{2}$$

where $c$ and $\eta$ are constants.

Lemma 1 has been proven in [13]. Using Lemma 1 we have

*Proposition 1:* Let $\rho_1(x,y) = \sqrt{x^2 + y^2}$, $\alpha_1 = D_i$, we have

$$Q_i(r,\theta) = \frac{2}{\pi D_i^2} \exp\left(-\frac{2r}{D_i}\right) \tag{3}$$

And let $\rho_2(x,y) = x^2 + y^{2\,\dagger}$, $\alpha_2 = D_i^2$, we have

$$Q_i(r,\theta) = \frac{1}{\pi D_i^2} \exp\left(-\frac{r^2}{D_i^2}\right) \tag{4}$$

where ($r$, $\theta$) is the polar coordinates, $D_i$ is a parameter related to the random environment, $r = \sqrt{x^2 + y^2}$, and $\theta$ is the azimuth angle.

Eq. (3) is proven in [11], and the proof of (4) is similar to (3), which is given in Appendix I. If the cell-side length of lattices is represented as $a$ in meters, we have the following

---

† $\rho_1$ and $\rho_2$ are the metrics corresponding to the trajectories of rays. The former can be regarded as the Euclidean distance, and the latter the square of the Euclidean distance.

*Proposition 2:* There exists an average distance among the closed clusters in the site percolation with a given *p*, denoted as $\bar{d}$, such that

$$\bar{d} = a/\sqrt{1-p} \tag{5}$$

**Proof**: In a 2D space, if the length and the width of lattice are with *N*, the total occupied planar area is $(Na)^2$. Since the closed site is occupied with probability 1-*p*, the number of closed site in this square is $N^2(1-p)$. Moreover, since the average distance among the closed clusters is $\bar{d}$, the area occupied by all the closed clusters is $N^2(1-p)\bar{d}^2$. Let $N^2(1-p)\bar{d}^2 = (Na)^2$, apparently $\bar{d}^2(1-p) = a^2$. As a result, Eq. (5) is immediately followed. Note also that, as $p \to 1, \bar{d} \to \infty;\ p \to 0, \bar{d} \to a$.

It is worth emphasizing that $\bar{d}$ is an important topological parameter of 2D site percolation lattices. Now let us denote the average distance traveled by a stochastic ray in *i* steps as $D_i$, and we further assume that

$$D_i = \bar{d}\, i^\beta, i \geq 1, \beta > 0 \tag{6}$$

where *i* means the number of re-radiations of a stochastic ray during its propagation and *β* is a parameter representing the anomalous level of stochastic rays. A similar formula to (6), $D_i = \alpha\, i^\beta, i \geq 1, \beta > 0$, is introduced in [8], where *α* is a variable depending on the probability *p*. In order to derive more rigorous and useful expression we substitute *α* for the invariant topological parameter of 2D site percolation in this paper. Substituting (5) and (6) into (3) and (4), the probability of stochastic rays at a special spatial location after undergoing *i* reflections is obtained immediately.

The distribution of particles obeying 2D random walks at site (*r*, *θ*) starting from the origin is as follows [14]

$$Q(r,\theta;t) = \frac{1}{4\pi Dt} e^{-r^2/(4Dt)} \tag{7}$$

where *t* is the time and *D* is the diffusion coefficient. After appropriate transformation of variables we have $Dt = \frac{1}{4}\bar{d}^2 i$ [14], where $\bar{d}$ is the step of random walks and *i* means the number of jumps. Since $D_i^2 = 4Dt$ [14] and $D_i = \bar{d}\, i^{1/2}$ [15], it is clear that (7) is the same as (4). Therefore if the probability of stochastic rays obeys (4), we called the obtained path loss model as random walk model; otherwise, if the probability of stochastic ray obeys (3), the path loss model as generic stochastic ray model. The parameter *β* represents the anomalous level of stochastic rays, which is related to the density of the lattice [8]. However, in order to obtain simpler analytic expressions of path loss, we only consider the cases of *β*=1/2 and 1. The former value corresponds to a moderately anomalous level, while the latter is associated with a random lattice channel with higher level of irregularity.

The probability distribution of stochastic rays for different models (random walk model and generic stochastic ray model with different $\beta$) is illustrated in Fig. 3, where the same channel geometrical parameters ($a = 20, p = 0.7, r = 150$) are considered. As can be seen from Fig. 3, the average number of stochastic rays undergoing less reflections increases from random walk model to generic stochastic ray models. Parallel trend is observed as $\beta$ increases from 1/2 to 1 for generic stochastic ray models.

## 3  Path loss models

Since $Q_i(r,\theta)$ is a pdf, we have $\iint_{r,\theta} Q_i(r,\theta)\,r\,\mathrm{d}r\,\mathrm{d}\theta = 1$. This means that if we consider the electromagnetic radiation in 2D plane as a probability space, then $Q_i(r,\theta)$ measures the energy diffusion in an ideal lossless propagation space. Furthermore, if we do not consider the effects of antennas, the mean received power in random lattice channels under non-line-of-sight (NLOS) propagation condition can be computed as follows

$$P(r,\theta) = P_T \sum_{i=1}^{\infty} 10^{-\frac{1}{10}\sum_{k=1}^{i} L_{ik}} Q_i(r,\theta), r > 1 \qquad (8)$$

where $P_T$ is the transmission power and $i$ represents the number of multiple bounces of a stochastic ray during the propagation process. $L_{ik}$ is the reflection loss in decibels due to the re-radiations from scatterers, which measures the energy loss of the $i$th MPC after the $k$th collision. Meanwhile $L_{ik}$ also measures the frequency dependence of channels, where the higher carrier frequency, the larger $L_{ik}$. This is consistent to the results of [16]. The condition $r > 1$ is imposed to fulfill the far-field scattering requirement. In a complicated propagation medium such as wireless multipath channels, it would be difficult to describe the values of $L_{ik}$ at all locations and at all times. Therefore, we characterize the medium statistically and assume that $L_{ik}$ can be modeled as a random variable.

In the following discussion, the abscissa component $\theta$ is omitted for simplicity. Moreover, because the received power is in the sense of statistical average, without loss of generality, we let $L_{ik} = L$, $\xi = L \ln 10 / 10$, $P_T = 1$. The mean received power in random lattice channels derived from (8) is given as follows

$$P(r) = \sum_{i=1}^{\infty} e^{-\xi i} Q_i(r) \qquad (9)$$

The path loss at site $r$, denoted as $PL(r)$, in decibels, is as follows
$$PL(r) = -P(r) \quad \text{(dB)} \qquad (10)$$

### 3.1  The random walk model
By replacing the summation operator in (9) with integration, we have

$$P(r) \approx \frac{1-p}{\pi a^2} \int_1^{\infty} x^{-1} \exp\left[-\xi x - \frac{(1-p)r^2}{a^2 x}\right] \mathrm{d}x \qquad (11)$$

Using [17, Eq. (3.478 4)]

$$P(r) \approx \frac{2(1-p)}{\pi a^2} K_0\left[2r\sqrt{\xi(1-p)}/a\right] \tag{12}$$

where $K_0(\cdot)$ is the modified Bessel function[17]. Using [17, Eq. (8.451 6)]

$$P(r) \approx \frac{(1-p)^{3/4} \xi^{-1/4}}{a\sqrt{\pi ar}} e^{-2r\sqrt{\xi(1-p)}/a}, \; 2r\sqrt{\xi(1-p)}/a \gg 1 \tag{13}$$

*3.2 The generic stochastic ray model*

When $\beta=1/2$, by replacing the summation operator in (9) with integration we have

$$P(r) \approx \frac{4(1-p)}{a^2\pi} \int_{a/\sqrt{1-p}}^{\infty} \frac{1}{y} \exp\left[-\frac{(1-p)\xi}{a^2} y^2 - \frac{2r}{y}\right] dy$$

$$\approx \frac{2(1-p)}{a^2\pi^{3/2}} G_{0,3}^{3,0}\left(\frac{\sqrt{(1-p)\xi} r}{a}, 2^{-1} \middle| \begin{array}{c} - \\ 0,0,\frac{1}{2} \end{array}\right) \tag{14}$$

where Meijer G-function [17] is used. Eq. (14) can be expressed in an approximate formula [11]:

$$P(r) \approx \frac{2}{ay_0}\sqrt{\frac{1-p}{3\pi\xi}} e^{-\frac{3r}{y_0}} \tag{15}$$

where $y_0 = \sqrt[3]{\frac{a^2 r}{(1-p)\xi}}$. The approximation error of (15) is discussed in [11], where we find that (15) is highly accurate for a very wide range of parameters.

When $\beta=1$, we also have

$$P(r) \approx \frac{2\sqrt{1-p}}{\pi a} \int_{a/\sqrt{1-p}}^{\infty} y^{-2} \exp\left[-\frac{\xi\sqrt{1-p} y}{a} - \frac{2r}{y}\right] dy$$

$$\approx \frac{2\sqrt{2\xi}}{\pi}\left(\frac{\sqrt{1-p}}{a}\right)^{3/2} r^{-1/2} K_1\left[2\sqrt{\frac{2\sqrt{1-p}\xi r}{a}}\right] \tag{16}$$

where [17, Eq. (3.478 4)] is used. Also using [17, Eq. (8.451 6)]

$$P(r) \approx \frac{(2\xi)^{1/4}}{\sqrt{\pi}}\left(\frac{\sqrt{1-p}}{a}\right)^{5/4} r^{-3/4} \exp\left(-2\sqrt{\frac{2\sqrt{1-p}\xi r}{a}}\right) \tag{17}$$

*3.3 Frequency dependence of the stochastic ray models*

A closer examination of Eqs. (13), (15) and (17) shows that there is an important factor $\exp(-\xi^\gamma)$ in the expression of the mean received power, where $\gamma$ is a constant. Thus, a larger $\xi$ would lead to a smaller value of $P(r)$. Furthermore, in an earlier discussion we have pointed out that the parameter $L_{ik}$ is the reflection loss in decibels due to the re-radiations from scatterers, which measures the energy loss of the $i$th MPC after the $k$th collision. As $\xi = L_{ik} \ln 10/10$ and a higher carrier frequency leads to a larger $L_{ik}$, $\xi$ can be regarded as a function of the carrier frequency. We can infer that the frequency dependence of the path loss models (Eqs. (13), (15) and (17)) is described by the parameter $\xi$. In the same propagation environment, the value of $\xi$ (or equivalently, $L_{ik}$) will be larger if the carrier frequency is increased.

## 4 Verification and discussions

In this section, we first investigate the determination of lattice parameters and the dependence of model outputs on the different parameters. Subsequently, we compare the results derived from Eqs. (13), (15) and (17) to the published experimental results under the NLOS conditions.

### *4.1 Determination of lattice parameters*

Three important lattice parameters should be determined: the cell-side length of lattices (*a*), the probability of each vertex to be open (*p*), and the reflection loss in decibels due to the re-radiations from scatterers (*L*). All of these values should correspond to the underlying propagation environment to be investigated. A simple rule is described as follows: *p* is the ratio of the occupation area of obstacles to the total area of considered region, *a* can be obtained from Eq. (5) if $\bar{d}$ is known, or it can be derived directly from the distribution of obstacles. The two parameters, *p* and *a*, are invariant for different models in the same propagation environment.

To determine the value of *L* is a more difficult task. It is unlikely to give exact values of *L* when we attempt to fit the various groups of measurement data, especially if the electrical properties of the scattering materials and the incident angles of different multipath components are not described in the relevant literature. Furthermore, the stochastic ray model is used to characterize an *average* propagation condition. Therefore, the value of *L* can only represent a mean reflection loss. There is a reasonable range of the value of *L* based on the physical appeal of radio propagation and generally, $2 \leq L \leq 10$ in decibels. I.e., *L*=3 dB indicates that there is 50% energy loss during each collision. The higher the carrier frequency, the larger the value of *L* [16].

The value of *L* in random walk model can be determined from the results of average propagation loss for radio paths obstructed by common building materials [16]. Since the average number of stochastic rays undergoing less reflections will change in different stochastic ray models as illustrated in Fig. 3, different values of *L* should be adopted for the different models (i.e., random walk model or generic stochastic ray model with different *β*) in order to ensure that the path loss in the same propagation scenario is similar. Following from Fig. 3, the value of *L* should increase from the random walk model to generic stochastic ray models and from *β*=1/2 to *β*=1. In general, the value of *L* in the random walk model should be 1~2 dB smaller than the generic stochastic ray model with *β*=1/2, and *L* in the latter should be 1~2 dB smaller than the generic stochastic ray model with *β*=1.

### *4.2 The effects of model parameters on the path loss*

This subsection investigates the sensitivity of the Eqs. (13), (15), (17) to different parameters, such as the choice of the metric, the values of *β*, and the lattice parameters. Under the same *a*, *p* and *L* (or *ξ*), and with a sufficiently large *r* (in order to guarantee the accuracy of the approximation formulas), the path loss under different models (the model of random walk and the models of generic stochastic ray) is plotted in Fig. 4, where $a = 20, p = 0.7, L = 3$. Fig. 4 demonstrates that the path loss is slightly sensitive to different stochastic-ray models. The path loss is more significant in the model of random walk. Nevertheless, similar trends of the path loss with respect to the propagation distance can be observed for all the three models.

Fig. 5 plots the path loss of the random walk model with different lattice parameters, where the variation of the parameters is 10%. (i.e., if $a = 20, p = 0.7, L = 3$, $\Delta a = 2, \Delta p = 0.07, \Delta L = 0.3$). Fig. 5 shows that the path loss is not very sensitive to different lattice parameters.

*4.3 Outdoor case*

The experimental data shown in [6] have been reproduced in Figs. 6(a)-6(c) and marked with circles. The measurements are taken in the "Prati" district of Rome, which is a classical dense urban area. A continuous wave at a frequency of 900 MHz is utilized. We assume that the interval distance *a*=20 m which simulates the average width of boulevards, the inoccupation probability *p*=0.7. Since the carrier frequency is 900 MHz, the loss due to reflection from scatterers is around 5.00 dB. We let the value of *L* be 3.5 in (13), 5.5 in (15) and 7.5 in (17), respectively, all in decibels. The path loss calculated from Eqs. (13), (15) and (17) are plotted in Figs. 6(a)-6(c). It can be seen that the theoretical results exhibit good agreement with the experimental data in all the three models. Furthermore, the root mean square (RMS) error can be used to evaluate the curve-fitting accuracy using the following measure

$$\sigma = \sqrt{\frac{1}{n}\sum_{i=1}^{n}[P_m(r_i) - P_t(r_i)]^2} \qquad (18)$$

where the subscripts *m* and *t* denote the measured data and the theoretical results obtained from (13), (15) or (17), respectively. The RMS errors are shown in Fig. 6. It can be seen that the model deviation is within a satisfactory range. Furthermore, if the values of *a*, *p* and *L* change in a small range, the RMS error will also change in a small range. This result is illustrated in Fig. 5. Therefore, the model prediction error is small if the values of *a*, *p* and *L* are chosen reasonably. Here we do not choose the parameters of (13), (15) or (17) by directly fitting with the experimental data. If we use the functions as (13), (15) or (17) to fit the data, we can obtain rather minor RMS errors.

*4.4 Indoor case*

The experimental data shown in [18] have been reproduced in Figs. 7(a)-7(c) and also marked with circles, which are calibrated at reference distance *r*=1.5 m. The measurements are taken in the laboratory environment at 60 GHz, where the multipath channel is a classical dense indoor propagation environment. Following the description of the measurement details in [18], we assume that the interval distance *a*=2 m, which simulates the average width between obstacles, the inoccupation probability *p*=0.82. Since the carrier frequency is 60 GHz, the loss due to the reflection from scatterers is larger and we let it be around 7.00 dB. We let the value of *L* be 6.0 in (13), 7.0 in (15) and 8.0 in (17), respectively, all in decibels. The path loss calculated from Eqs. (13), (15) and (17) are plotted in Figs. 7(a)-7(c), where the RMS errors are depicted. It can be seen that the model deviation is also within a satisfactory range.

*4.5 Discussions*

Path loss models in random lattice channels can be obtained directly from the method of stochastic rays. The approximation accuracy of (13), (15) and (17) is very high, which can be seen from [11] and the derivation procedures of corresponding closed-form expressions. Therefore, the analytical framework presented in this paper is very useful in the study of propagation characteristics for random lattice channels. Moreover, we can see from the expressions of (13), (15) and (17) that the mean received power is a function of $r^{-1/2}e^{-r}$, $r^{-1/3}e^{-r^{2/3}}$, and $r^{-3/4}e^{-r^{1/2}}$, respectively. The similar mathematical structure is also derived by the transport theory in [4, 10]. But the results in this paper are easy to use in practical electromagnetic engineering. Because the exponential function is rigorous monotonic, the path loss described by (17) has the flattest trend in propagation environment. This tendency is also can be seen from Fig. (3). Nevertheless, a large amount of measurement data is required to provide further insight into the selection of proper model parameters.

## 5  Conclusions

In this paper, the path loss models for random lattice channels have been derived. Closed-form expressions of the probability that a ray undergoes certain reflections at a specified location have been derived under the Euclidean distance. The pdf of stochastic rays obtained from maximum entropy approaches under the special constraint is the same as the distribution of particles obeying random walks. Therefore the stochastic rays can be divided into two classes: stochastic rays obeying random walks and generic stochastic rays. The paper investigates three kinds of models: the model of random walks and the models of generic stochastic rays in the moderately anomalous channel ($\beta=1/2$) and the channel with higher anomalous level ($\beta=1$). It has been demonstrated that the path loss can be represented as a closed-form function of three important channel parameters, namely, the average inter-obstacle distance, the non-occupation probability of obstacles, and the reflection loss due to scattering elements. All of these parameters can be derived directly from the geometrical and electrical properties of obstacles in propagation environment under consideration. The path loss prediction is robust to different lattice parameters. The model results have a good agreement with empirical data for both outdoor and indoor propagation scenarios.

**Acknowledgments**
This work is supported by the NSFC No. 60572024. The authors would like to thank the anonymous reviewers for their valuable comments.

**Appendix I**
Proof of Eq. (4)
Let $\lambda = e^\eta$, we have $ce^{\eta \rho_i(x,y)} = c\lambda^{\rho_i(x,y)}$. Let $Q_i(x,y) = Q_0 \lambda^{x^2+y^2}$. Since $\iint_{x,y} Q_i(x,y)\mathrm{d}x\mathrm{d}y = 1$, we have

$$\iint_{x,y} Q_0 \lambda^{x^2+y^2} \mathrm{d}x\mathrm{d}y = \iint_{r,\theta} Q_0 \lambda^{r^2} r\mathrm{d}r\mathrm{d}\theta$$

$$= Q_0 \int_0^{2\pi} \mathrm{d}\theta \int_0^\infty r\lambda^{r^2} \mathrm{d}r = 2\pi Q_0 \int_0^\infty r\lambda^{r^2} \mathrm{d}r.$$

Let $J_0 = \int_0^\infty r\lambda^{r^2} \mathrm{d}r$, we have $J_0 = -\dfrac{1}{2\ln\lambda}$, which requires $\lambda < 1$, otherwise the integral expression will be divergent. Then

$$\lambda = \exp(-\pi Q_0), \quad Q_i(r,\theta) = Q_0 \exp(-\pi Q_0 r^2),$$

$$\int_{\theta=0}^{2\pi}\int_{r=0}^\infty r^2 Q_0 \exp(-\pi Q_0 r^2) r \mathrm{d}r \mathrm{d}\theta$$

$$= \int_0^{2\pi} Q_0 \mathrm{d}\theta \int_0^\infty r^3 \exp(-\pi Q_0 r^2) \mathrm{d}r$$

$$= 2\pi Q_0 \int_0^\infty r^3 \exp(-\pi Q_0 r^2) \mathrm{d}r$$

$$= \frac{1}{\pi Q_0} \int_0^\infty (\pi Q_0 r^2) \exp(-\pi Q_0 r^2) \mathrm{d}(\pi Q_0 r^2)$$

$$= \frac{1}{\pi Q_0} \int_0^\infty z \mathrm{e}^{-z} \mathrm{d}z = \frac{1}{\pi Q_0} \Gamma(2)$$

$$= \frac{1}{\pi Q_0} = D_i^2,$$

$Q_0 = \dfrac{1}{\pi D_i^2}$, therefore Eq. (4) is derived immediately.

References

1. Bertoni, H. L.: 'Radio Propagation for Modern Wireless Systems'(NJ: Prentice Hall, 1999), Chap. 2
2. Franceschetti, G., Marano, S., and Palmieri, F.: 'Propagation without wave equation, toward an urban area model'. *IEEE Trans. Antennas Propagat.*, 1999, 47, (9), pp. 1393-1404
3. URL: http://www.diei.unipg.it/MEL/mel_abstracts/radiopropagation_and_em_compatibility.html
4. Ullmo, D., and Baranger, H.U.: 'Wireless propagation in buildings: a statistical scattering approach'. *IEEE Trans. Vehicular Technol.*, 1999, 47, (9), pp. 947-955
5. Moustakas, A.L., Baranger, H.U., Balents, L., Sengupta, A.M., and Simon, S.H.: 'Communication through a Diffusive Medium: Coherence and Capacity'. *Science*, 2000, 287, 287-290
6. Franceschetti, M., Bruck, J., and Schulman, L.: 'A random walk model of wave propagation'. *IEEE Trans. Antennas Propagat.*, 2004, 52, (5), pp. 1304-1317
7. Franceschetti, M.: 'Stochastic rays pulse propagation'. *IEEE Trans. Antennas Propagat.*, 2004, 52, (10), pp. 2742-2752
8. Marano, S., and Franceschetti, M.: 'Ray propagation in a random lattice: a maximum

entropy, anomalous diffusion process'. *IEEE Trans. Antennas Propagat.*, 2005, 53, (6), pp. 1888-1896

9   Hu, L.-Q., and Zhu, H.: 'Applications of stochastic bridge methods for UWB wireless channel'. *J. of Electronics & Information Technology*, 2006, 28, (10), pp. 1846-1850

10  Janaswamy, R.: 'An indoor pathloss model at 60 GHz based on transport theory'. IEEE Antennas and Wireless Propagat. Lett., 2006, 5, pp. 58-60

11  Hu, L.-Q., and Zhu, H.: 'No-wave approaches and its application to received power of radio wave propagation'. 2006 China-Japan Joint Microwave Conference, Chengdu, China, 2006, 2, pp. 605-609

12  Grimmett, G.: 'Percolation'(NY: Springer-Verlag, 1989), Chap. 5

13  Cover, T. M., and Thomas, J. A.: 'Elements of Information Theory'(NY: Wiley, 1991), Chap. 11

14  Zhang, Q.: 'Statistical Mechanics'(Beijing: Science press, 2004), Chap. 10

15  Holland, B.K.: 'What are the Chances?: Voodoo Deaths, Office Gossip, and Other Adventures in Probability' (Baltimore: The Johns Hopkins Univ. Pr., 2002), Chap. 6

16  Rappaport, T.S.: 'Wireless Communications Principles and Practice'(NY: Prentice-Hall, 1996), Chap. 3

17  Gradshteyn, I.S., and Ryzhik, I.M.: 'Table of Integrals, Series, and Products' (NY: Academic, 6th ed., Jeffrey, A., Ed., 2000), Chap. 3 & 8

18  Moraitis, N., and Constantinou, P.: 'Indoor channel measurements and characterization at 60 GHz for wireless local area network applications'. *IEEE Trans. Antennas Propag.*, 2004, 52, (12), pp. 3180-3189

**Figure captions:**

Fig. 1 *Site percolation lattices, where the interval a=2, and unoccupied probability p=0.3, 0.5, 0.7, 0.9, respectively*

Fig. 2 *Stochastic rays propagating in random lattice channels, where Tx is a transmitter, Rx is a receiver*

Fig. 3 *The probability distribution of stochastic rays under the different models with the same channel geometrical parameters ( $a=20, p=0.7, r=150$ )*

Fig. 4 *The path loss under different stochastic rays but with the same lattice parameters*

Fig. 5 *The path loss of the model of random walk with different lattice parameters, where $a=20, p=0.7, L=3$ and $\Delta a=2, \Delta p=0.07, \Delta L=0.3$*

Fig. 6 *The path loss calculated from theoretical models versus measurements in outdoor environment [6], a=20 m, p=0.7; solid line is theoretical results, the circles are the experimental data*
*a   calculated from (13)*
*b   calculated from (15)*
*c   calculated from (17)*

Fig. 7 *The path loss calculated from theoretical models versus measurements in indoor environment [18], a=2 m, p=0.82; solid line is theoretical results, the circles are the experimental data*
*a   calculated from (13)*
*b   calculated from (15)*
*c   calculated from (17)*

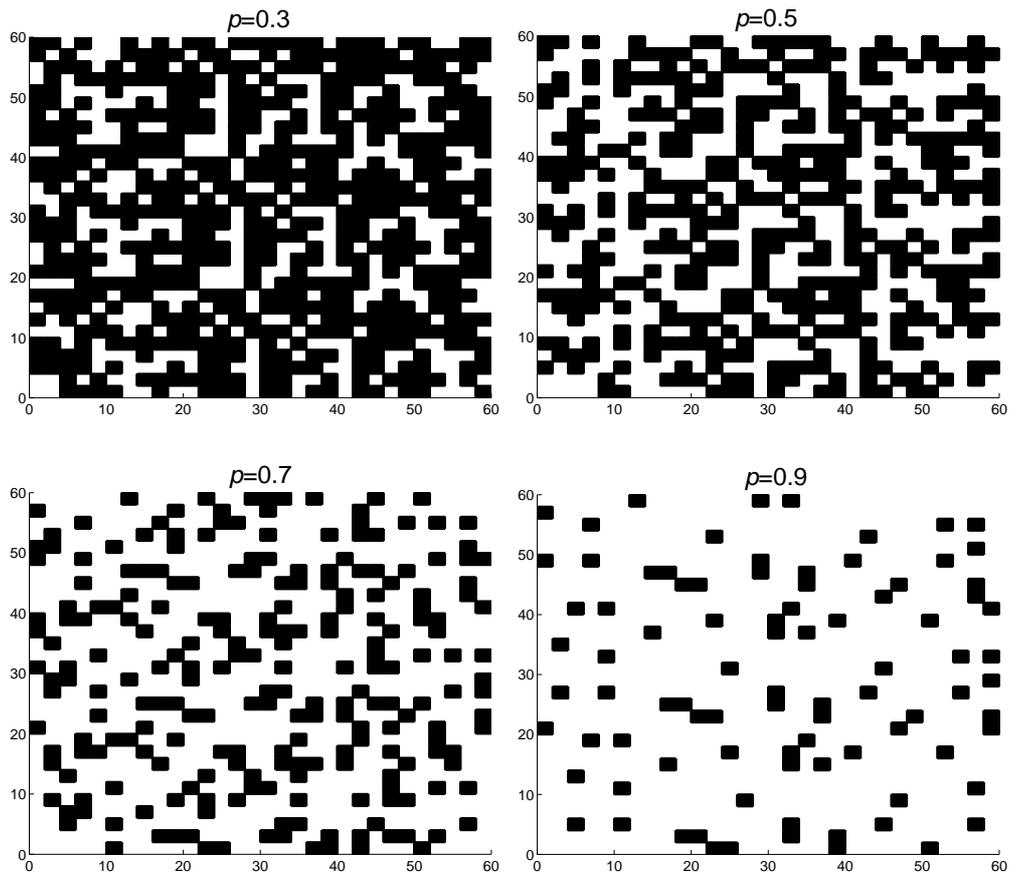

Fig. 1

L.-Q. Hu, *et al*.

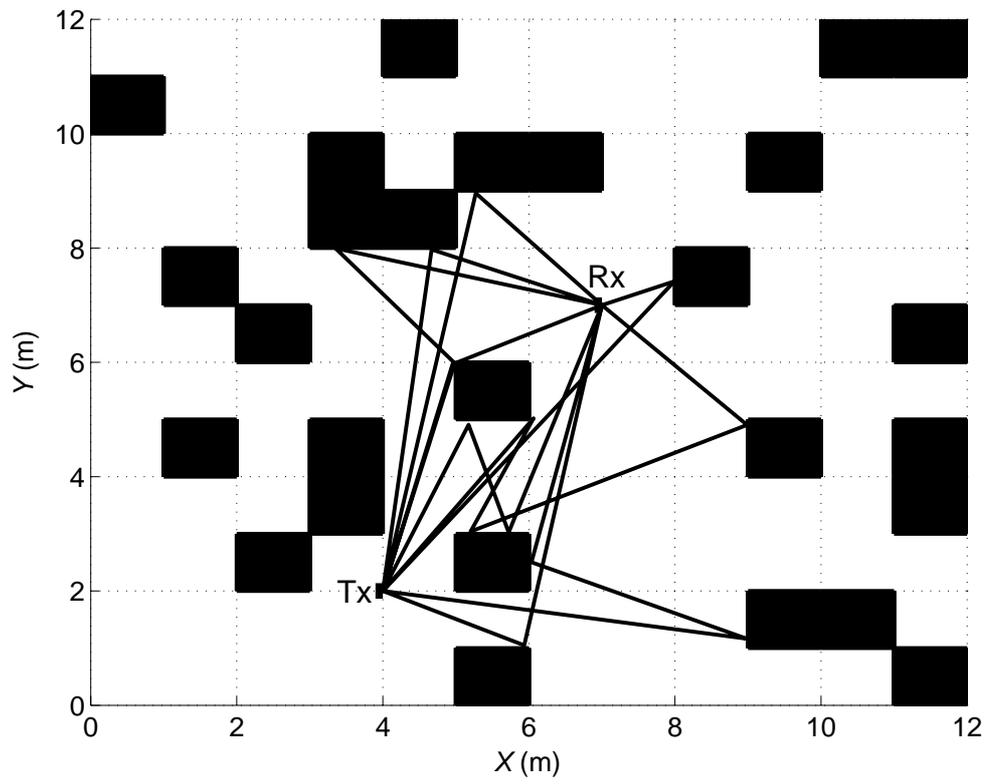

Fig. 2

L.-Q. Hu, *et al*.

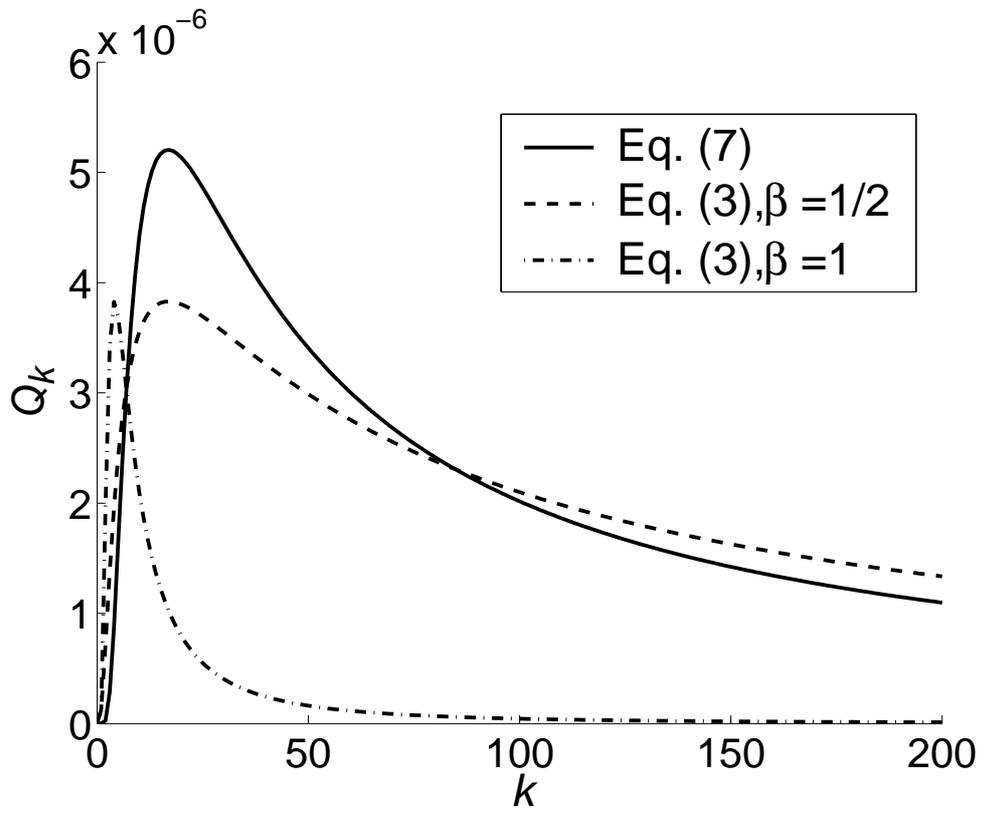

Fig. 3

L.-Q. Hu, *et al*.

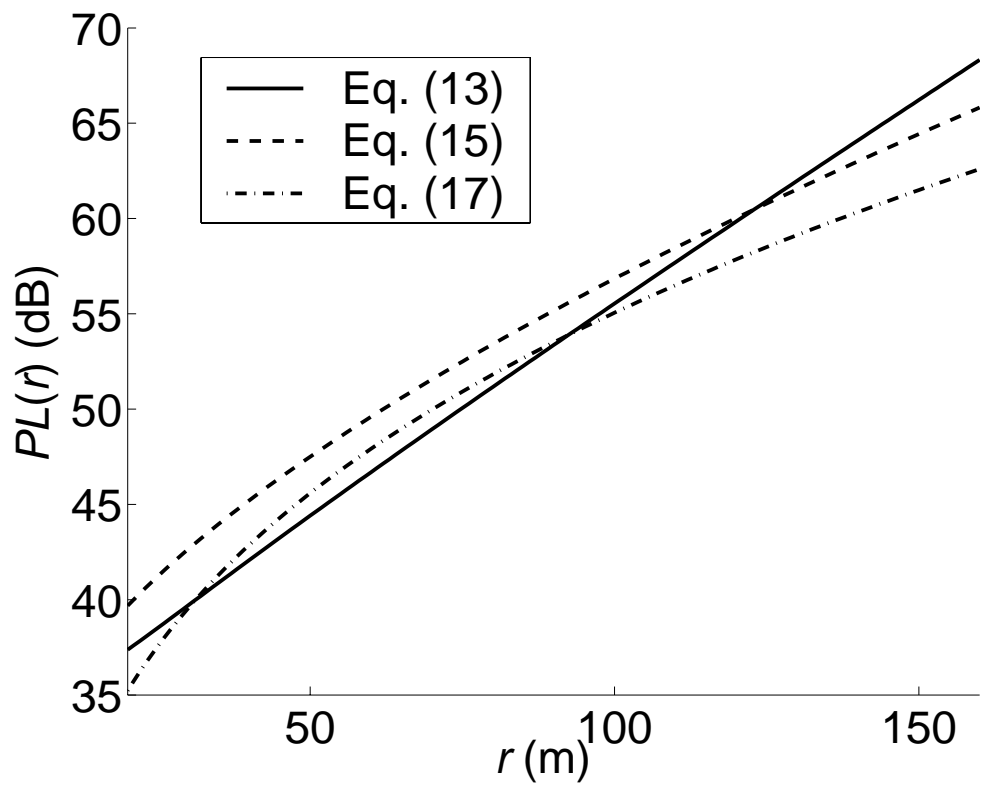

Fig. 4

L.-Q. Hu, *et al*.

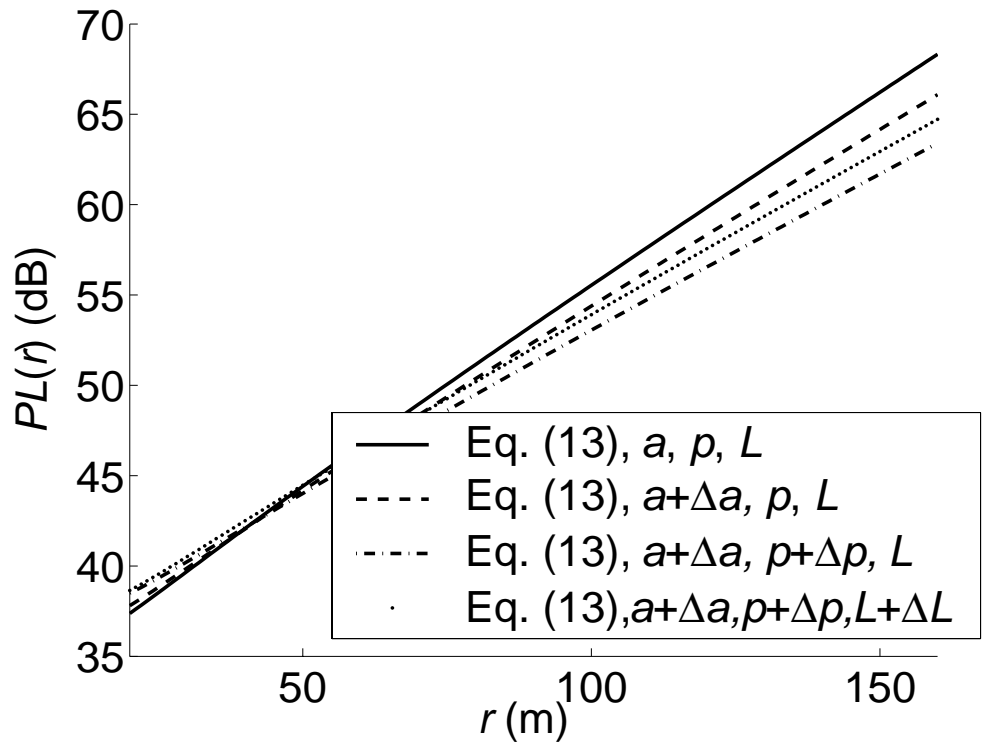

Fig. 5

L.-Q. Hu, *et al*.

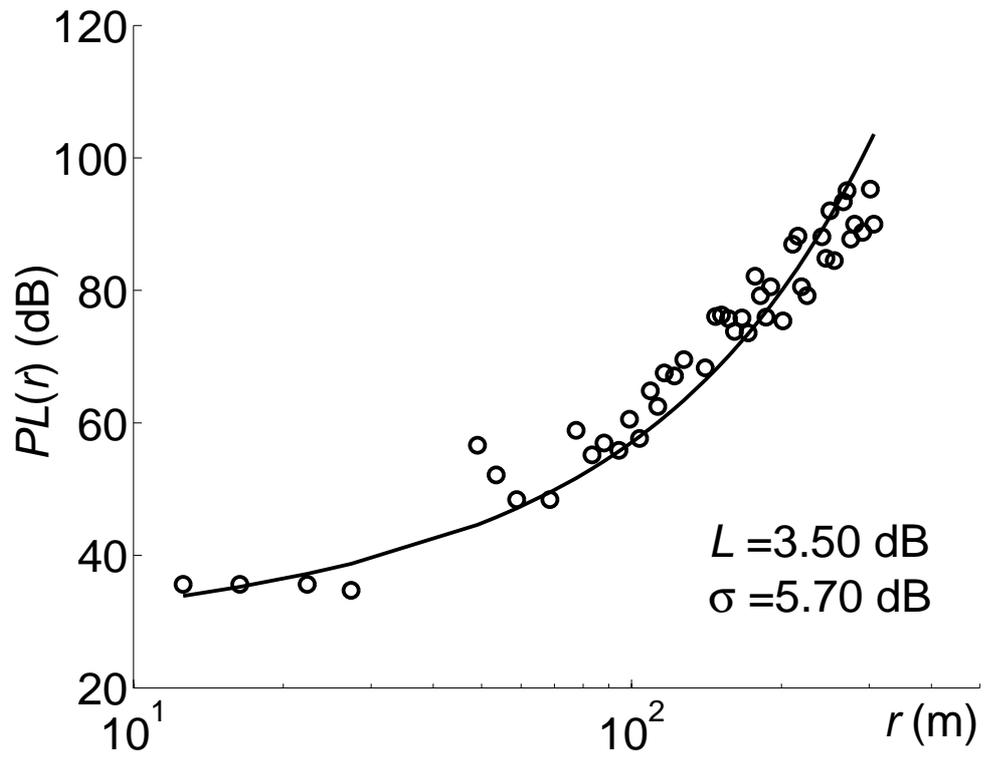

Fig. 6 (a)

L.-Q. Hu, *et al*.

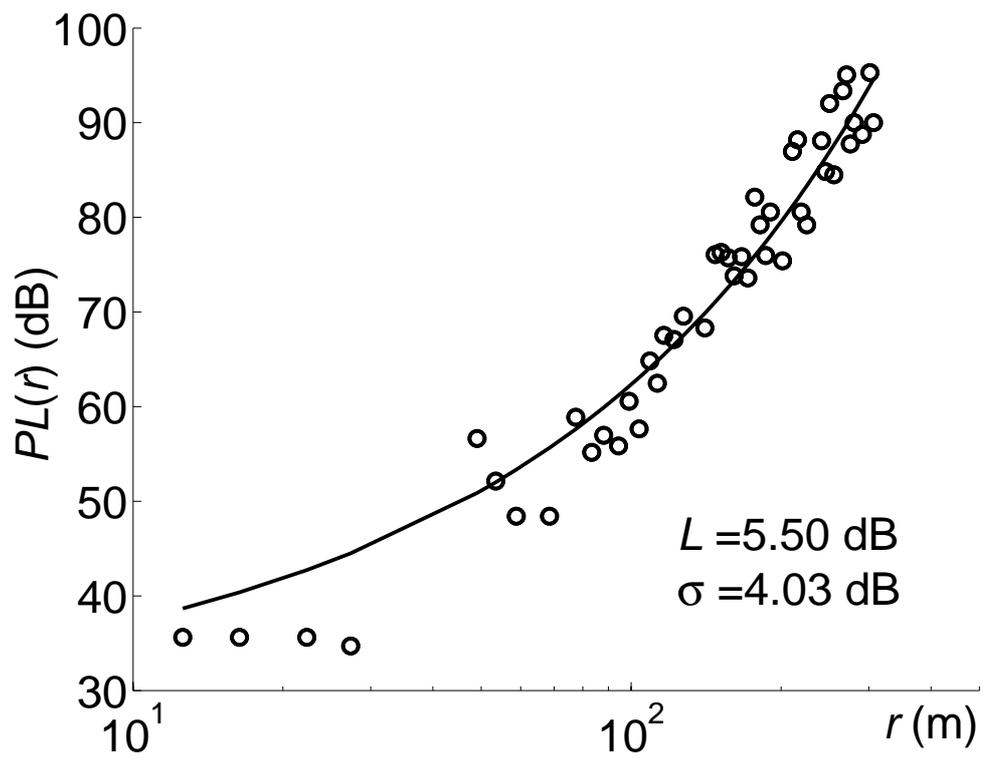

Fig. 6 (b)

L.-Q. Hu, *et al*.

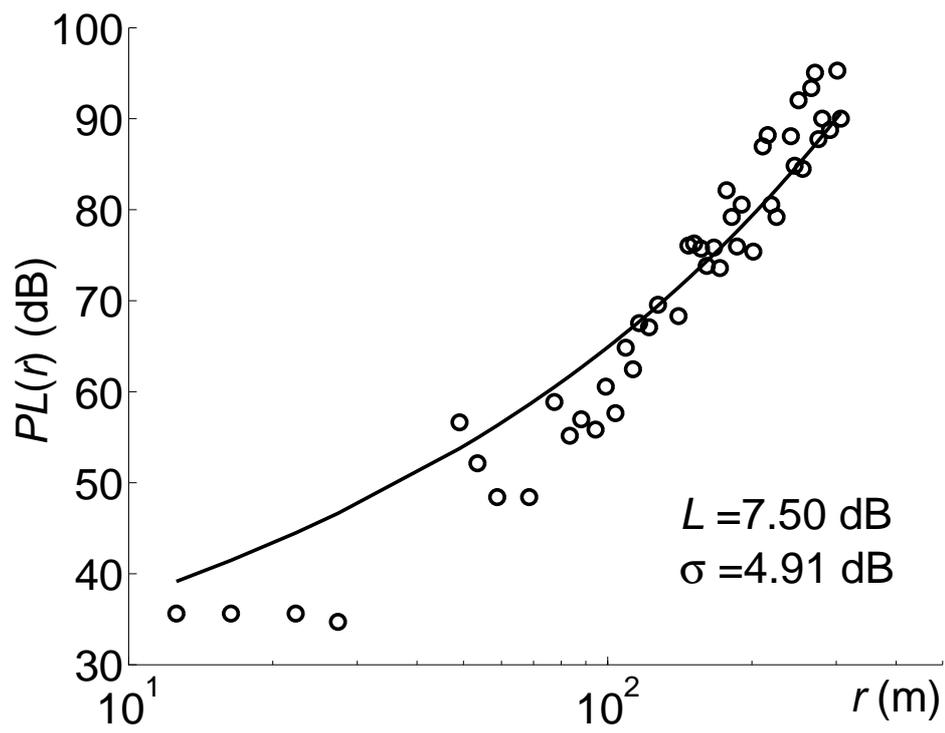

Fig. 6 (c)

L.-Q. Hu, *et al*.

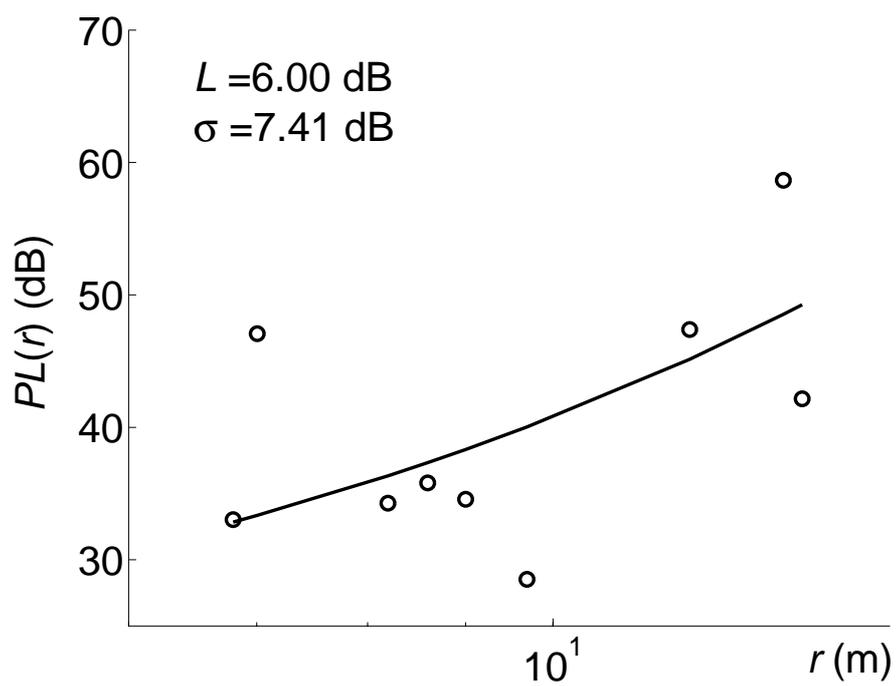

Fig. 7 (a)

L.-Q. Hu, *et al*.

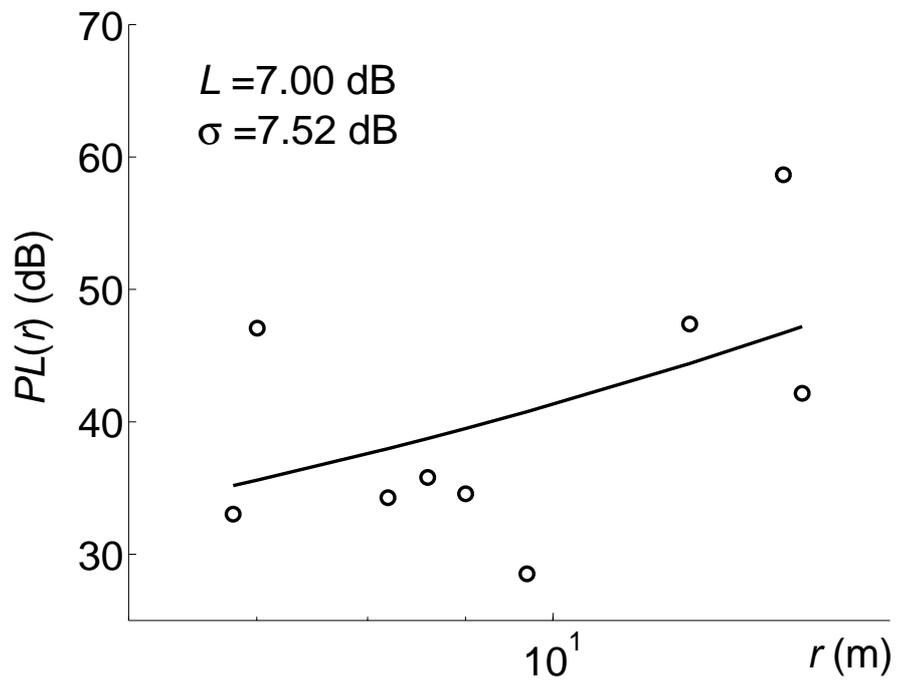

Fig. 7 (b)

L.-Q. Hu, *et al*.

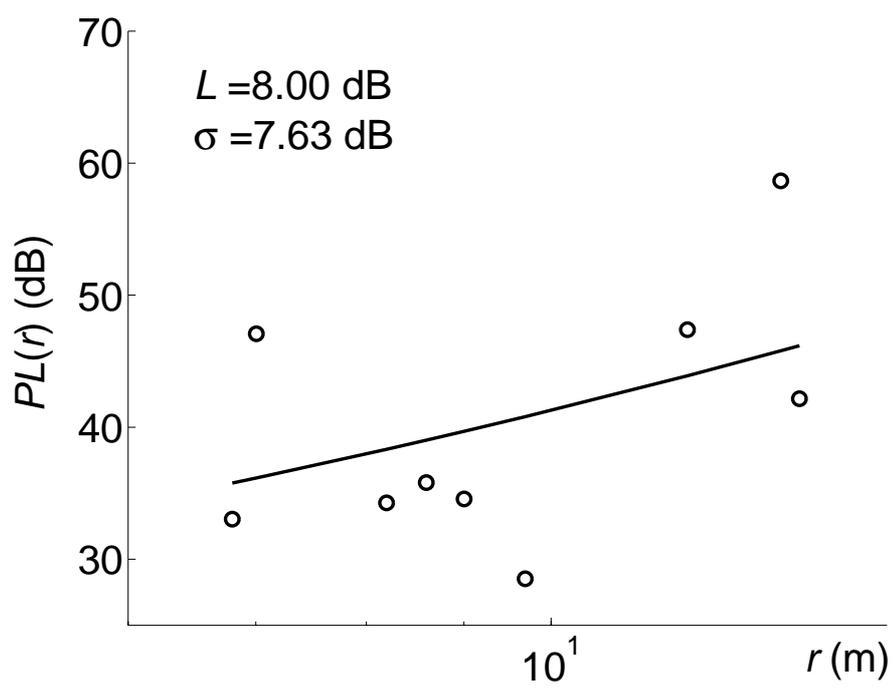

Fig. 7 (c)

L.-Q. Hu, *et al*.